\documentstyle[buckow2]{article} 
\input epsf

\def\beq{\begin{equation}} 
\def\eeq{\end{equation}}

\def\etal{{\it et al.}~}

\def\eg{{\it e.g.},~}

\def\3he{$^3$He}
\def\4he{$^4$He}
\def\6li{$^6$Li}
\def\7li{$^7$Li}

\newcommand{\obh}{$\Omega_{\rm B}h^2$}

\def\la{\mathrel{\mathpalette\fun <}}
\def\ga{\mathrel{\mathpalette\fun >}}
\def\fun#1#2{\lower3.6pt\vbox{\baselineskip0pt\lineskip.9pt
  \ialign{$\mathsurround=0pt#1\hfil##\hfil$\crcr#2\crcr\sim\crcr}}}
 
\begin {document} 

\begin{center}

\vspace*{1cm}

{\Large\bf The Baryon Density Through The (Cosmological) Ages 
 
\vskip 1truecm 

Gary Steigman}

\large

\vskip 5truemm  

Departments of Physics and Astronomy\\ 
The Ohio State University\\ 
174 West 18th Avenue\\
Columbus, OH 43210, USA\\ 

\vskip 1truecm 
 
{\bf Abstract}

\end{center}

\large

The light element abundances probe the baryon density of the universe 
within a few minutes of the Big Bang.  Of these relics from the earliest 
universe, deuterium is the baryometer of choice.  By comparing its 
primordial abundance inferred from observations with that predicted by 
Big Bang Nucleosynthesis (BBN), the early universe baryon density is 
derived.  This is then compared to independent estimates of the baryon 
density several hundred thousand years after the Big Bang and at present, 
more than 10 billion years later.  The excellent agreement among these 
values represents an impressive confirmation of the standard model of 
cosmology.

\large 
 
\section{Introduction}

In the new, precision era of cosmology, {\bf redundancy} will play an 
increasingly important role, permitting multiple, independent, tests 
of and constraints on competing cosmological models, and providing a 
window on systematic errors which can impede progress or send us off 
in unprofitable directions.  To illustrate the value of this approach, 
here the baryon density of the Universe is tracked from the first few 
minutes (as revealed by BBN), through the first few hundred thousand 
years later (as coded in the fluctuation spectrum of the Cosmic Microwave 
Background -- CMB), and up to the present epoch, approximately 10 Gyr 
after the expansion began.  Theory suggests and terrestrial experiments 
confirm that the baryon number is preserved throughout these epochs, so 
that the number of baryons ($\equiv$~nucleons) in a comoving volume {\it 
should} be unchanged from BBN to today.  As a surrogate for tracking a 
comoving volume, the nucleon density may be compared to the density of 
CMB relic photons.  Except for the additional photons produced when 
e$^{\pm}$ pairs annihilate, the number of photons in any comoving 
volume is also preserved.  As a result, the baryon density may be traced 
throughout the evolution of the universe utilizing the nucleon-to-photon 
ratio $\eta \equiv n_{\rm N}/n_{\gamma}$.  Since the temperature of the 
CMB fixes the present number density of relic photons, the fraction of 
the critical mass/energy density in baryons (nucleons) today ($\Omega_{\rm 
B} \equiv \rho_{\rm B}/\rho_{crit}$) is directly related to $\eta$ by, 
$\eta_{10} \equiv 10^{10}\eta = 274\Omega_{\rm B}h^{2}$, where the Hubble 
parameter is H$_{0} \equiv 100h$~kms$^{-1}$Mpc$^{-1}$.  According to the 
HST Key Project, $h = 0.72 \pm 0.08$ \cite{HST}.

For several decades now the best constraints on $\eta$ have come from
the comparison of the predictions of BBN with the primordial abundances 
of the relic nuclides D, \3he, \4he, and \7li, as inferred from 
observational data.  New data, of similar accuracy, which will soon be 
available, will enable quantitative probes of the baryon density at later 
epochs in the evolution of the universe.  Indeed, recent CMB data has 
very nearly achieved this goal.  Although a comparable level of precision 
is currently lacking for the present universe estimates, the data do permit 
comparisons of independent estimates of $\eta$ (or \obh) at three widely 
separated eras in the evolution of the universe.  

\section{The Baryon Density During The First Few Minutes}\label{sec:BBN}

During the first few minutes in the evolution of the universe the density
and temperature were sufficiently high for nuclear reactions to occur.  As 
the universe expanded, cooled, and became more dilute, the universal nuclear 
reactor ceased to create or destroy nuclei.  The abundances of the light 
nuclei formed during this epoch are determined by the competition between 
the time available (the universal expansion rate) and the density of the 
reactants: neutrons and protons.   The abundances of D, \3he, and \7li are 
``rate limited", being determined by the competition between the nuclear 
production/destruction rates and the universal expansion rate.  As such, 
they are sensitive to the nucleon density and have the potential to serve 
as ``baryometers".  
\begin{figure}
  \centering
  \epsfysize=3.5truein
  \epsfbox{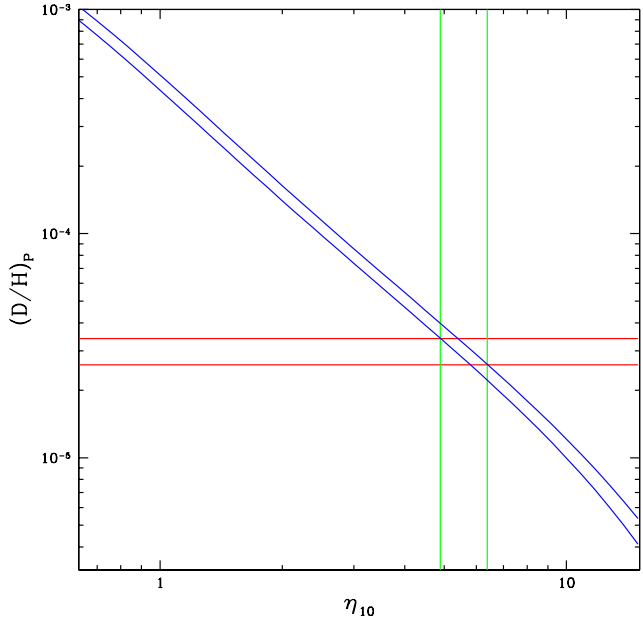}
  \caption{The band stretching from upper left to lower right is the
BBN-predicted deuterium abundance.  The horizontal band is the observational 
estimate of the primordial abundance (see the text).  The vertical band 
provides an estimate of the BBN-derived baryon density.}\label{fig:dvseta}
\end{figure}

Of the light, relic nuclides whose primordial abundances may probe the 
baryon density, deuterium is the {\bf baryometer of choice}.  Its predicted 
primordial abundance varies sigificantly with the nucleon density (D/H 
$\propto \eta^{-1.6}$) (see Fig. 1).  As a result, a primordial abundance 
known to, say, 10\%, determines the baryon density ($\eta$) to $\sim 6\%$; 
truly precision cosmology!  Equally important, BBN is the only astrophysical 
site where an ``interesting" abundance of deuterium may be produced (D/H 
$\ga 10^{-5}$) \cite{els}; the relic abundance is not enhanced by post-BBN 
production.  Furthermore, as primordial gas is cycled through stars, 
deuterium is completely destroyed.  Because of the small binding energy of 
the deuteron, this destruction occurs during pre-main sequence evolution, 
when stars are fully mixed.  As a result, the abundance of deuterium 
will only have decreased (or, remained close to its primordial value) since 
BBN.
\begin{figure}
  \centering
  \epsfysize=3.5truein
  \epsfbox{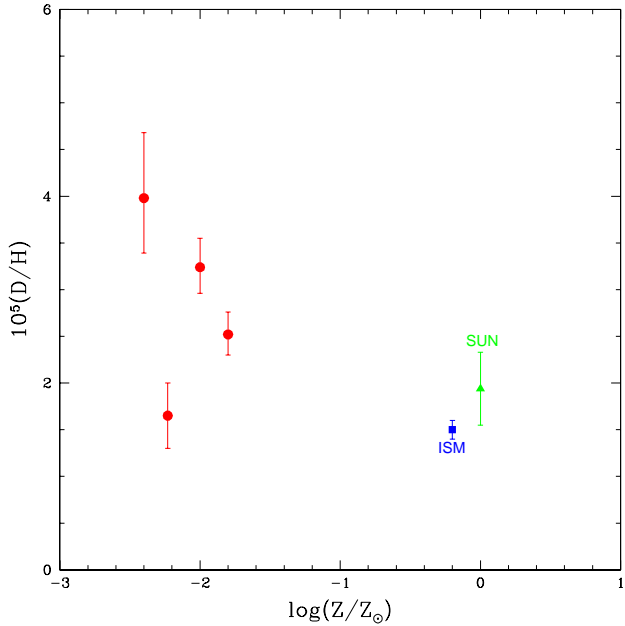}
  \caption{The deuterium abundance by number relative to hydrogen derived 
from observations of high-z, low-Z QSO absorption line systems as a function 
of metallicity (relative to solar).  Also shown are the deuterium abundances 
inferred for the local insterstellar medium \cite{linsky} and for the presolar 
nebula \cite{gg}.}\label{fig:dvsz1}
\end{figure}

After some false starts, as of early 2001 there were three high-redshift 
(z), low-metallicity (Z) QSO absorption line systems where deuterium had 
been reliably detected \cite{bt98a,bt98b,O'M}.  Earlier observations 
\cite{hid} of a system with very high D/H are widely agreed to have had
insufficient velocity data to rule out contamination of the deuterium 
absorption by a hydrogen interloper \cite{antihid}.  More recently, 
detection of deuterium has been claimed for another QSO absorption line 
system \cite{DOd1,DOd2}.  However, the large variations in the inferred 
D/H, due to the uncertain velocity structure, render this line of sight 
inappropriate for determining the primordial deuterium abundance.  An 
apparently more reliable, recent determination along a different line of 
sight yields a surprisingly small D/H \cite{PB}.  These current data, along 
with the D/H values for the local interstellar medium \cite{linsky} and the 
pre-solar nebula \cite{gg} are shown in Figure 2 as a function of 
metallicity.  At low metallicity there should be a deuterium ``plateau", 
whose absence, so far, is notable -- and very puzzling.  Clearly more data 
is needed.  For the purpose of deriving the early universe baryon density, 
I will rely on the deuterium abundance proposed by O'Meara \etal \cite{O'M}: 
(D/H)$_{\rm BBN} \equiv 3.0 \pm 0.4 \times 10 ^{-5}$.

From a careful comparison between the BBN predicted abundance and the 
adopted primordial value, the baryon density when the universe is less 
than a half hour old is determined (see Fig. 1): $\eta_{10}({\rm BBN}) 
= 5.6 \pm 0.5 ~(\Omega_{\rm B}h^{2} = 0.020 \pm 0.002)$.  This is truly 
a ``precision" determination; only time will tell if it is accurate.  
The likelihood distribution of this BBN-derived baryon density is the 
curve labelled ``BBN" in Figure \ref{fig:lik3}.     

\section{The Baryon Density A Few Hundred Thousand Years Later}\label{sec:cmb}

\begin{figure}
  \centering
  \epsfysize=3.83truein
  \epsfbox{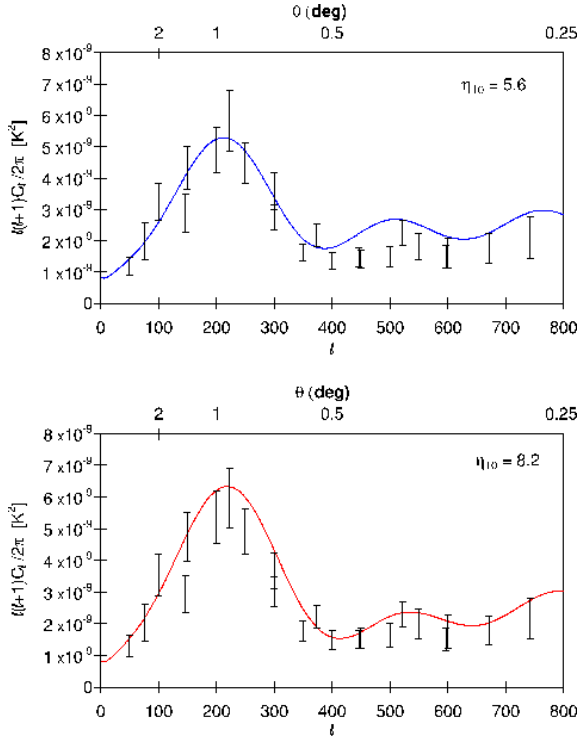}
  \caption{The CMB angular fluctuation spectra for two models which 
differ only in the adopted baryon density.  The BBN inferred baryon 
density is shown in the upper panel and, for comparison, a higher 
baryon density model is shown in the lower panel.  The data are 
from the ``old" BOOMERANG and MAXIMA-1 observations; see the text 
for references.}\label{fig:cmb}
\end{figure}

The early universe is radiation dominated and the role of ``ordinary" matter 
(baryons) is subdominant.  As the universe expands and cools, the density 
in non-relativistic matter becomes relatively more important, eventually 
dominating after a few hundred thousand years.  At this stage perturbations 
begin growing under the influence of gravity and, on scales determined by 
the relative density of baryons, oscillations in the baryon-photon fluid 
develop.  At redshift $\sim 1100$, when the electron-proton plasma combines 
to form neutral hydrogen, the CMB photons are freed to propagate thoughout 
the universe (``last scattering").  These CMB photons preserve the record 
of the baryon-photon oscillations through very small temperature fluctuations 
in their spectrum.  These fluctuations, or anisotropies, have been detected 
by the newest generation of experiments, beginning with COBE \cite{cobe} and 
continuing with the exciting early results from BOOMERANG \cite{boom1,boom2} 
and MAXIMA-1 \cite{max1}, providing a tool for constraining the baryon 
density at last scattering.  In Figure~\ref{fig:cmb} the status quo ante is 
shown.  The relative heights of the odd and even peaks in the CMB angular
fluctuation spectrum depend on the baryon density and these early BOOMERANG 
and MAXIMA-1 data favored a ``high" baryon density (compare the ``BBN case", 
$\eta_{10} = 5.6$, in the upper panel of Fig. 3 with that for a baryon density 
some 50\% higher shown in the lower panel).  These data posed a challenge 
to the consistency of the standard models of cosmology and particle physics, 
suggesting that the baryon number may have changed (increased) since BBN.  
Subsequently, new data from the DASI experiment \cite{DASI}, along with 
the revised and expanded BOOMERANG \cite{BOOM} and MAXIMA \cite{MAX} data 
appeared.  The new and revised data eliminated the challenge posed by the 
older results.  

Although the extraction of cosmological parameters from the CMB anisotropy 
spectra can be very dependent on the priors adopted in the analyses (see \eg
\cite{kssw}), the baryon density inferred when the universe was a few hundred 
thousand years old is robust.  For example, we (Kneller \etal \cite{kssw}) 
find, $\eta_{10}({\rm CMB}) = 6.0 \pm 0.6 ~(\Omega_{\rm B}h^{2} = 0.022 \pm 
0.002)$.  The likelihood distribution for this CMB-determined baryon abundance
is shown in Figure \ref{fig:lik3} by the curve labelled ``CMB".  The excellent 
agreement between the two independent estimates, BBN at a few minutes, and 
CMB at a few hundred thousand years, represents a spectacular success for 
the standard model of cosmology and illustrates the great potential for 
future precision tests of cosmology.

\section{The Baryon Density At 10 Gyr}\label{sec:sn1a}

There are a variety of approaches to measuring the baryon density today, or 
during the very recent past.  Many of these depend on assumptions concerning 
the relation between mass and light or, they require the adoption of a 
specific model for the growth of structure.  The approach utilized here 
attempts to avoid such model-dependent assumptions.  Instead, the data from 
the SNIa magnitude-redshift surveys \cite{scp1,hiz,scp2}, is 
used along with the {\it assumption} of a flat universe (this latter receives 
strong support from the newest CMB data \cite{DASI,BOOM,MAX}) 
to pin down the total matter density ($\Omega_{\rm M}$).  This is then 
combined with an estimate of the universal baryon {\bf fraction} ($f_{\rm 
B} \equiv \Omega_{\rm B}/\Omega_{\rm M}$) derived from studies of the X-ray 
emission from clusters of galaxies.  For more details on this approach, see 
\cite{shf} and \cite{swz}.  

\begin{figure}
  \centering
  \epsfysize=3.0truein
  \epsfbox{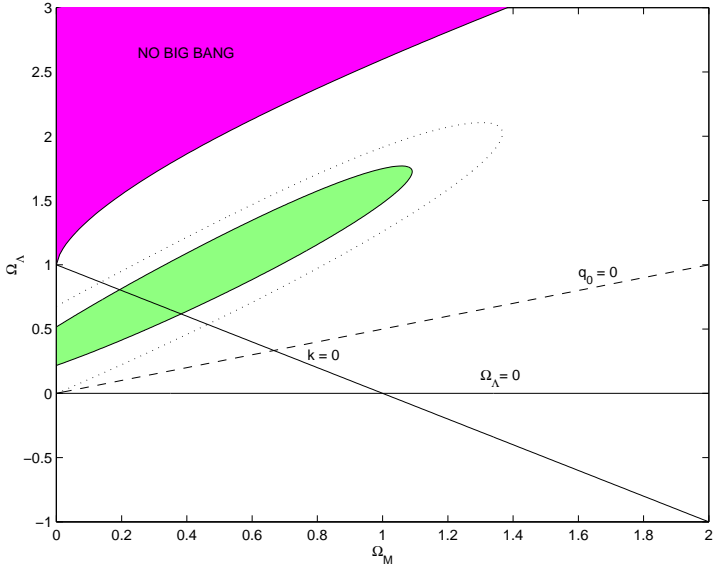}
  \caption{The 68\% (solid) and 95\% (dotted) contours in the 
$\Omega_{\Lambda} - \Omega_{\rm M}$ plane allowed by the SNIa
magnitude-redshift data (see the text for references).  
Geometrically flat models lie along the diagonal line labelled 
$k = 0$.}\label{fig:sn1a}
\end{figure}

In Figure~\ref{fig:sn1a} are shown the SNIa-constrained 68\% and 
95\% contours in the $\Omega_{\Lambda} - \Omega_{\rm M}$ plane.
The expansion of the universe is currently accelerating for those
models which lie above the (dashed) q$_{0} = 0$ line.  The $k = 0$
line is for a ``flat" (zero 3-space curvature) universe.  As shown 
in \cite{swz}, adopting the assumption of flatness and assuming the 
validity of the SNIa data, leads to a reasonably accurate ($\sim 25\%$)
estimate of the present matter density: $\Omega_{\rm M}(\rm{SNIa;Flat}) 
= 0.28 ^{+0.08} _{-0.07}$.

As the largest collapsed objects, rich clusters of galaxies provide an 
ideal probe of the baryon {\bf fraction} in the present/recent universe.  
Observations of the X-ray emission from clusters of galaxies permit 
constraints on the hot gas content of such clusters which, when corrected 
for the baryons in stars (but, unfortunately, not for any dark baryons!), 
may be used to estimate (or bound) $f_{\rm B}$.  From observations of the 
Sunyaev-Zeldovich effect in X-ray clusters, the hot gas fraction may be 
constrained \cite{grego} and used to estimate $f_{\rm B}$ \cite{skz}.  
The combination of $f_{\rm B}$ and $\Omega_{\rm M}$ is then used to derive 
a present-universe ($t_{0} \approx 10$~Gyr; $z \la 1$) baryon density: 
$\eta_{10}({\rm SNIa;Flat}) = 5.1 ^{+1.8} _{-1.4} ~(\Omega_{\rm B}h^{2} 
= 0.019 ^{+0.007} _{-0.005})$.  In Figure~\ref{fig:lik3} the corresponding 
likelihood distribution for the present universe baryon density is shown 
labelled by``SNIa".  Although the uncertainties are largest for this 
present-universe value, it is in excellent agreement with the other, 
independent estimates.

\section{Summary -- Concordance}\label{sec:summ}

The abundances of the relic nuclides produced during BBN reflect the 
baryon density present during the first few minutes in the evolution 
of the universe.  Of these relics from the early universe, deuterium 
is the baryometer of choice.  Although more data is to be desired,
the current data permit reasonably constrained estimates of the
abundance of primordial deuterium, leading to a tight constraint on
the early universe baryon-to-photon ratio $\eta_{10}({\rm BBN}) = 
5.6 \pm 0.5$ ($\Omega_{\rm B}h^{2} = 0.020 \pm 0.002$).

\begin{figure}
  \centering
  \epsfysize=3.0truein
  \epsfbox{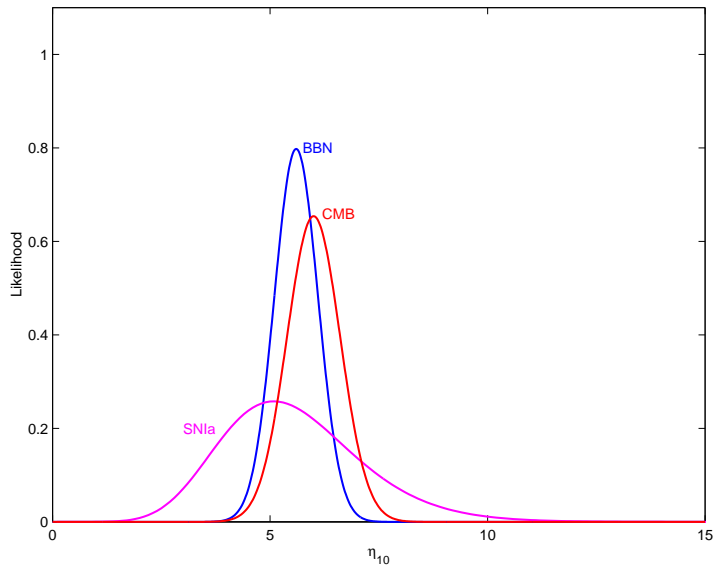}
  \caption{The likelihood distributions (normalized to equal areas 
under the curves) for the baryon-to-photon ratios derived from BBN, 
from the CMB, and for the present universe (SNIa).
}\label{fig:lik3}
\end{figure}

Several hundred thousand years later, when the universe became transparent 
to the CMB radiation, the baryon density was imprinted on the temperature
fluctuation spectrum.  In determining the baryon density, the current CMB 
data have a precision approaching that of BBN: $\eta_{10}({\rm CMB}) = 6.0 
\pm 0.6$ ($\Omega_{\rm B}h^{2} = 0.022 \pm 0.002$).  The excellent agreement 
between the BBN and CMB values (see Fig.~\ref{fig:lik3}) provides strong 
support for the standard model of cosmology.

In the present universe most baryons are dark ($\eta ({\rm LUM}) \ll 
\eta ({\rm BBN}) \approx \eta ({\rm CMB})$), so that estimates of the 
baryon density some 10 billion years after the expansion began are more 
uncertain and, often model-dependent.  In \S\,\ref{sec:sn1a} an estimate 
of the total matter density ($\Omega_{\rm M}$) derived from the SNIa 
magnitude-redshift data was combined with the assumption of a flat 
universe to derive $\eta_{10}({\rm SNIa;Flat}) = 5.1 ^{+1.8} _{-1.4}$ 
($\Omega_{\rm B}h^{2} = 0.019 ^{+0.007} _{-0.005}$).  Although this 
estimate is of lower statistical accuracy than those from BBN or the 
CMB, it is in agreement with them (see Fig.~\ref{fig:lik3}).  Note 
that {\it if} the mass of dark baryons in clusters were similar to 
the stellar mass, the present-universe baryon density estimate would 
increase by $\sim 10\%$, bringing it into essentially perfect overlap 
with the BBN and CMB values.  

Increasingly precise observational data have permitted us to track the 
baryon density from the big bang to the present.  At widely separated 
epochs, from the first few minutes, through the first few hundred 
thousand years, to the present, $\sim$ ten billion year old universe, 
a consistent value for the baryon abundance is revealed.  This remarkable 
concordance of the standard, hot, big bang cosmological model is strikingly 
revealed by the overlapping likelihood distributions for the universal 
baryon abundances shown in Figure~\ref{fig:lik3}.  

\begin{center}
{\bf Acknowledgments}
\end{center}
I wish to thank the Alexander von Humboldt Foundation and its extremely
efficient staff for having organized such an exciting and informative
symposium.  The research described here is supported by the U.S. D.O.E.
through grant DE-FG02-91ER-40690.



\begin{thebibliography}{77} 

\bibitem{HST}
W. L. Freedman, {\it et al.}, ApJ {\bf 553} (2001) 47.

\bibitem{els}
R. Epstein, J. Lattimer, \& D. N. Schramm, Nature
{\bf 263} (1976) 198.

\bibitem{bt98a}
S. Burles \& D. Tytler, ApJ {\bf 499} (1998a) 699.

\bibitem{bt98b}
S. Burles \& D. Tytler, ApJ {\bf 507} (1998b) 732.

\bibitem{O'M}
J. M. O'Meara, {\it et al.}, ApJ {\bf 552} (2001) 718.

\bibitem{hid}
J. K. Webb, R. F. Carswell, K. M. Lanzetta, R. Ferlet, 
M. Lemoine, A. Vidal-Madjar, \& D. V. Bowen, Nature {\bf 388} 
(1997) 250.

\bibitem{antihid}
D. Kirkman {\it et al.}, ApJ {\bf 559} (2001) 23.

\bibitem{DOd1} 
S. D'Odorico, M. Dessauges-Zavadsky, \& P. Molaro,  
A\&A {\bf 338} (2001) L1.
 
\bibitem{DOd2}
S. A. Levshakov, M. Dessauges-Zavadsky, S. D'Odorico, 
\& P. Molaro, ApJ {\bf 565} (2002) 696. 

\bibitem{PB}
M. Pettini, \& D. V. Bowen, ApJ {\bf 560} (2001) 41.

\bibitem{linsky}
J. L. Linsky, \& B. E. Wood, Proceedings of IAU Symposium 
198, The Light Elements and Their Evolution (L. da Silva, 
M. Spite, and J. R. Medeiros eds.; ASP Conference Series), 
(2000) p. 141.

\bibitem{gg}
G. Gloeckler \& J. Geiss, Proceedings of IAU Symposium 
198, The Light Elements and Their Evolution (L. da Silva, 
M. Spite, and J. R. Medeiros eds.; ASP Conference Series), 
(2000) p. 224.

\bibitem{cobe}
C. L. Bennett, {\it et al.}, ApJ {\bf 464} (1996) L1.

\bibitem{boom1}
P. de Bernardis, {\it et al.}, Nature {\bf 404} (2000) 955.

\bibitem{boom2}
A. E. Lange, {\it et al.}, Phys. Rev. {\bf D 63} (2001) 042001.

\bibitem{max1}
S. Hanany, {\it et al.}, ApJ {\bf 545} (2000) L5.

\bibitem{DASI}
N. W. Halverson, {\it et al.,} preprint (2001) [astro-ph/0104489].

\bibitem{BOOM}
C. B. Netterfield, {\it et al.}, preprint (2001) [astro-ph/0104460].
 
\bibitem{MAX}
A. T. Lee, {\it et al.}, ApJ {\bf 561} (2001) L1.

\bibitem{kssw}
J. P. Kneller, R. J. Scherrer, G. Steigman, \& T. P. Walker, 
Phys. Rev. {\bf D 64} (2001) 123506.

\bibitem{scp1}
S. Perlmutter, {\it et al.}, ApJ {\bf 483} (1997) 565.

\bibitem{hiz}
B. P. Schmidt, {\it et al.}, ApJ {\bf 507} (1998) 46.

\bibitem{scp2}
S. Perlmutter, {\it et al.}, ApJ {\bf 517} (1999) 565.

\bibitem{shf}
G. Steigman, N. Hata, \& J. E. Felten, ApJ {\bf 510} (1999) 564.

\bibitem{swz}
G. Steigman, T. P. Walker, \& A. Zentner, preprint (2000) 
[astro-ph/0012149].

\bibitem{grego}
L. Grego, {\it et al.}, ApJ {\bf 552} (2001) 2.

\bibitem{skz}
G. Steigman, J. P. Kneller, \& A. Zentner, Revista Mexicana 
de Astronomia y Astrofisica {\bf 12} (2002) 265.

\end{thebibliography}
\end{document}